# Design of a W-band High-PAE Class A&AB Power Amplifier in 150nm GaAs Technology

Jun Yan Lee[a,b], Duo Wu[b], Xuanrui Guo[c,b], Mohammad Mahdi Ariannejad[b], Mohammad Arif Sobhan Bhuiyan[b,*] and Mahdi H. Miraz[d,e,f,*]

[a]School of Electronic, Information and Electrical Engineering, Shanghai Jiao Tong University, Shanghai, China;

[b]Department of Electrical and Electronic Engineering, Xiamen University Malaysia, Sepang, Malaysia;

[c]Department of Microelectronics, Delft University of Technology, Delft, Netherlands;

[d]School of Computing and Data Science, Xiamen University Malaysia, Sepang, Malaysia;

[e]Faculty of Arts, Science and Technology, Wrexham University, UK;

[f]Faculty of Computing, Engineering and Science, University of South Wales, Swansea, UK, e-mail: m.miraz@ieee.org.

Nanometer scale power amplifiers (PA) at sub-THz suffer from severe parasitic effects that lead to experience limited maximum frequency and reduced power performance at the device transceiver front end. The integrated circuits researchers proposed different PA design architecture combinations at scaled down technologies to overcome these limitations. Although the designs meet the minimum requirements, the power added efficiency (PAE) of PA is still quite low. In this paper, a W-band single-ended common-source (CS) and cascode integrated 3-stage 2-way PA design is proposed. The design integrated different key design methodologies to mitigate the parasitic; such as combined Class AB and Class A stages for gain-boosting and efficiency enhancement, Wilkinson power combiner for higher output power, linearity, and bandwidth, and transmission line (TL)-based wide band matching network for better inter-stage matching and compact size. The proposed PA design is validated using UMS 150-nm GaAs pHEMT using advanced design system (ADS) simulator. The results show that the proposed PA achieved a gain of 20.1 dB, an output power of 17.2 dBm, a PAE of 33 % and a 21

GHz bandwidth at 90 GHz Sub-THz band. The PA layout consumes only 5.66 × 2.51 mm² die space including pads. Our proposed PA design will boost the research on sub-THz integrated circuits research and will smooth the wide spread adoption of 6G in near future.

**1. Introduction**

The evolution of wireless communication systems has enabled to deploy the fifth-generation (5G) communication systems in practice to get benefit from their capabilities of high transmission rate and low latency [1]. However, the 5G communication system suffers from transmission rate and bandwidth limitations for the next frontier virtual communications applications such as holographic applications, virtual mode communication of AR/VR, digital twin, satellite IoT with global coverage, autonomous driving, and unmanned aerial vehicles [2]. Therefore, driving forward the sixth generation (6G) communication system is essential to further expand the service of the communication network. A power amplifier (PA), located at the end of the transmitter side, is one of the important as well as core modules that enable the signals to be transmitted efficiently as shown in Figure 1. The core objective of a PA is to ensure the desired signal amplification depending on its gain, the transmission distance coverage depending on the output power, the power consumption depending on the efficiency, the transmission quality depending on the linearity, and the transmission quantity based on the bandwidth. A typical PA contains a driver stage and a power stage along with necessary bias and impedance matching circuits [3], while more details about the 2 stages of PA will be explained in section below. The architecture of these modules determines the overall performance of a transceiver at a particular frequency band.

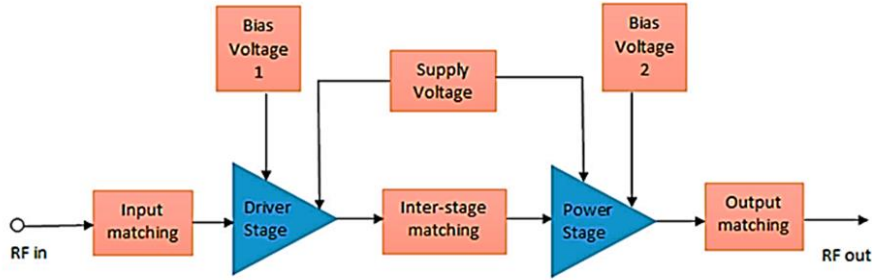

Figure 1. Block diagram of a basic PA modules.

The shifting from 5G to 6G communication, devices at sub-THz frequency bands introduces several disadvantages because the speed of the transistor does not significantly change from the lower millimeter-wave (mmW) region to a higher frequency region [4]. PA needs to meet high power requirements at low supply voltages in order to execute a better performance tradeoff. It is important to ensure high linearity to reduce heat dissipation and maintain high efficiency. Besides, designing a PA that operates at sub-THz frequency is also very challenging with the limited technology specifications currently available. Therefore, the PA designs have been tailored with different technology processes, such as InP HBT [5], InP HEMT [6], GaN [7], GaAs mHEMT [8], GaAs pHEMT [9], CMOS [10], and SiGe BiCMOS [11], to look into the design performance and optimization methods.

The literature review on designing 6G communication devices using InP, GaN, GaAs, and CMOS technologies, however, reveals that the efficiencies of PAs are still generally lower than 10 % [12-17]. Also, popular technology such as CMOS suffers from other limitations such as high output conductance, transit frequency/maximum oscillation frequency $f_t/f_{max}$, velocity saturation etc., as comprehensively presented in several articles [18-21]. On the other hand, GaAs technology shows an optimum performance at THz and sub-THz frequency bands as it exhibits a higher electron mobility which have a faster carrier movement with high power density for handling high power while providing better efficiency [22].

Besides, the literature review unveils various architectures and techniques reliably adopted for the sub-THz PA designs which includes different amplifier type, power combining and splitting for higher saturated output power [10, 13, 14, 16], multi-stage configurations for gain boosting [6-11, 14, 16, 17], neutralization method for high linearity and efficiency [10, 17], etc. Specifically, PA stages are connected in cascade [10], cascode [7, 8, 13, 17] , pseudo-differential [17], or differential [10] manner with PA common gate (CG) [11], common source (CS) [10], or common emitter (CE) [14] using different biasing and feedback techniques.

In this paper, therefore, a 150 nm GaAs pHEMT technology-based W-band PA architecture is proposed to mitigate the above design challenges. It includes:

(1) a capacitive biased cascode power stage that is able to precharge the biasing capacitor by parasitic voltage for self-biasing.

(2) low loci $\lambda$/4 wideband matching.

(3) square topology-based power combining technique for circuit compactness.

This paper is organized as follows: section 2 introduces the design methodology and method of designed PA; section 3 shows the designed PA's results and comparison with state-of-the-arts; and section 4 concludes the paper.

**2. Design Methodology**

A conventional PA consists of various blocks e.g., the input matching network, the driver stage, inter-stage matching network, power stage, output matching network, and biasing network. A PA design needs to be restructured for better performance at sub-THz bands. The block diagram of the proposed PA for 90 GHz transceiver application is illustrated in Figure 2. The PA consist of three-stages including two driver stages followed by a single power stage for better power performance. The design strategy of proposed multistage configuration adopts 2-way power division and combination method as well as transmission line (TL) based impedance matching techniques. Figure 3 shows the PA layout design with overall dimension of $5.66 \times 2.51$ mm$^2$ including pads. The following subsections provide the details of the design of the proposed PA:

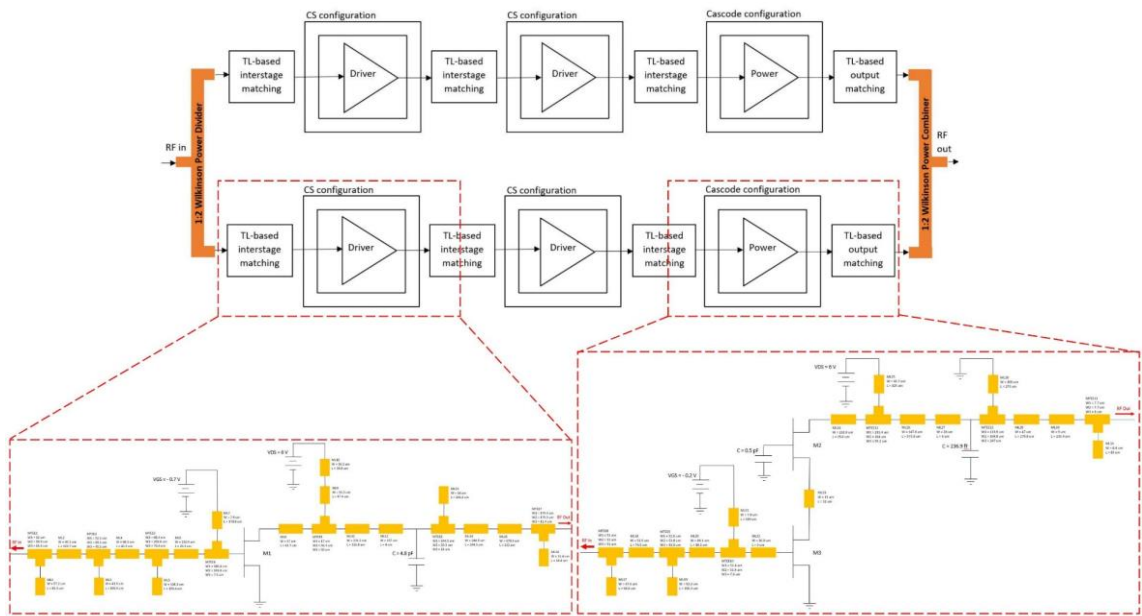

Figure 2. Block diagram of proposed PA and detailed schematic of CS and cascode stages.

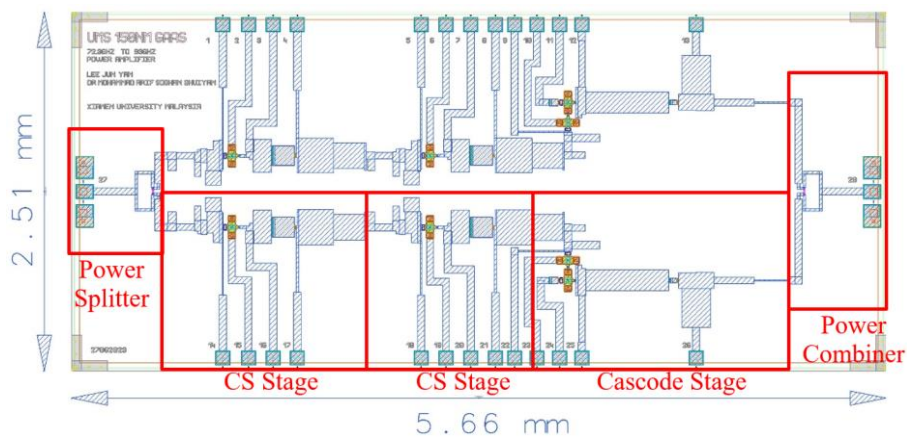

Figure 3. Layout design of the proposed PA.

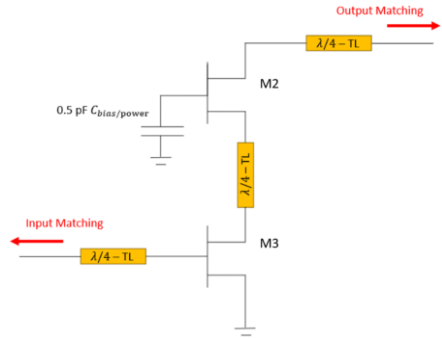

Figure 4. Schematic of capacitive biasing in power stage.

*2.1 Multistage PA*

A single stage PA cannot meet the power requirements. In order to boost the gain of the PA, a two stage driver stages followed by a single power stage design has been proposed in this research. The first and the second common-source (CS) stages are configured in Class-AB that combines the benefit of both class-A and class-B with its high efficiency and linearity properties to drive the PA in multi-stage design with minimized parasitic performance degradation. The transistor gate width was set to its minimum, i.e., 50 um with only 2 fingers to reduce the parasitic effect from the series inductance at the gate that will degrade the matching capability. Moreover, the $V_{ds}$ was set to 6 V and the gates, $V_{gs}$ were biased at -0.7 V based on the Direct-Current Current-Voltage (DC-IV) simulation.

The power stage is located at the final stage and is designed by adopting the cascode topology with its capability to deliver a high power amplification to the load. The cascode stage also serves for a high output impedance to maintain both the stability and the linearity. The cascode stage is configured in a Class-A architecture to utilize its high linearity and gain properties. A 0.5 pF capacitor was used as a biasing capacitor to overcome the excess parasitic voltage and maintain the stability of the power stage of the PA, as described in Figure 4. At high frequency operation, parasitic is the main cause of the heat dissipation and low efficiency. This biasing capacitor can take the advantage from the parasitic voltage for self-biasing precharge. Eventually, the PA stage will have higher efficiency and output power. The $V_{ds}$ was set to 6 V and the gate, $V_{gs}$ was biased at -0.2 V based on the DC-IV simulation.

*2.2 2-Way Power Divider*

The power combining and dividing method is a common method to improve the PA output power and efficiency [23-25]. The performance of the saturated power and efficiency will be degraded due to high operational frequency at down-scaled technologies. Therefore, a Wilkinson's 1: 2 power divider and a 2:1 power combiner using microstrip lines (ML) was designed to boost the output power and efficiency requirements, it is best for 3-db power enhancement. The Wilkinson's power divider has the properties of matched, isolated, and reciprocal which makes is the best among compare to T-junction divider and resistive divider. The power divider is designed on a $\lambda/4$ transformer to allow the split ports have a common port and a resistor to enable fully isolation at the center frequency of 90 GHz. The power divider also has flexibility to design in different layouts without degrading the performance, while a square topology-based power divider provides a smaller dimension [24].

*2.3 Wide-band Matching*

Impedance matching is always an important step to ensure minimum mismatch between each stage to minimize the degradation of overall gain, output power, efficiency and linearity performance. With the available source and load impedance results from the load pull analysis. The source impedances of CS stage and cascode stage are 7.547 + j5.71 and 10.968 + j5.003, respectively. The load impedance of CS stage and cascode stage are 13.347 + j24.349 and 41.672 - j15.218, respectively. The matching networks for the CS driver stage and the cascode power stage are designed using ML-based stub using wideband matching network with a matching loci stay near to Q ≈ 1 [26]. The matching network is also designed as $\lambda/4$ transformer-based TL due to its s-parameter stabilization capability and able to improve the inter-stage mismatch issue [27]. The matched circuits of the driver stage and the cascode power stage have been further optimized to meet the expected results operating at W-band as shown in Figure 2.

## 3. Simulation Result

A high-power UMS 150 nm GaAs pHEMT process technology was adopted to design and validate the proposed PA architecture using advanced design system (ADS) simulator. In this study, S-parameter analysis, harmonic balance analysis, output power and efficiency are evaluated. Moreover, Monte-Carlo analysis is also performed to verify the reliability of the PA performance. In this study, the input power is swept from 0 dBm to 20 dBm and the operating temperature of the PA circuit performance evaluation was set to 300K.

The S-parameter analysis of the proposed 2-way 3 stages PA are illustrated in Figure 5. It shows that the PA achieves an S21 gain of 20.054 dB at 90 GHz frequency and a 3-dB bandwidth of approximately 20 GHz (from 72.3 GHz to 92.95 GHz) covering 90 GHz resonant frequency. The Figure 6 (a) exhibits the linearity analysis which shows that the output power is increasing with the input power and getting saturated at 17.4 dBm for 90 GHz due to the influence of junction capacitances of the transistors used. At 85 GHz, the output power also increases linearly and saturated at 19.4 dBm. The output power at 95 GHz started to be unstable as there is a slightly drop during saturation due to the frequency is close to the maximum operating frequency of the transistor which performance is unstable. This also explains the Figure 6 (b) shows a higher PAE (48.7 %) operating at 85 GHz and lower PAE (26 %) at 95 GHz, while the maximum power added efficiency (PAE) for the PA is 33 % at 8 dBm input power when operating at 90 GHz. At higher input power, the PAE drastically decrease due to the transistor limitations, particularly when operating at high frequency that causes high parasitic and energy losses.

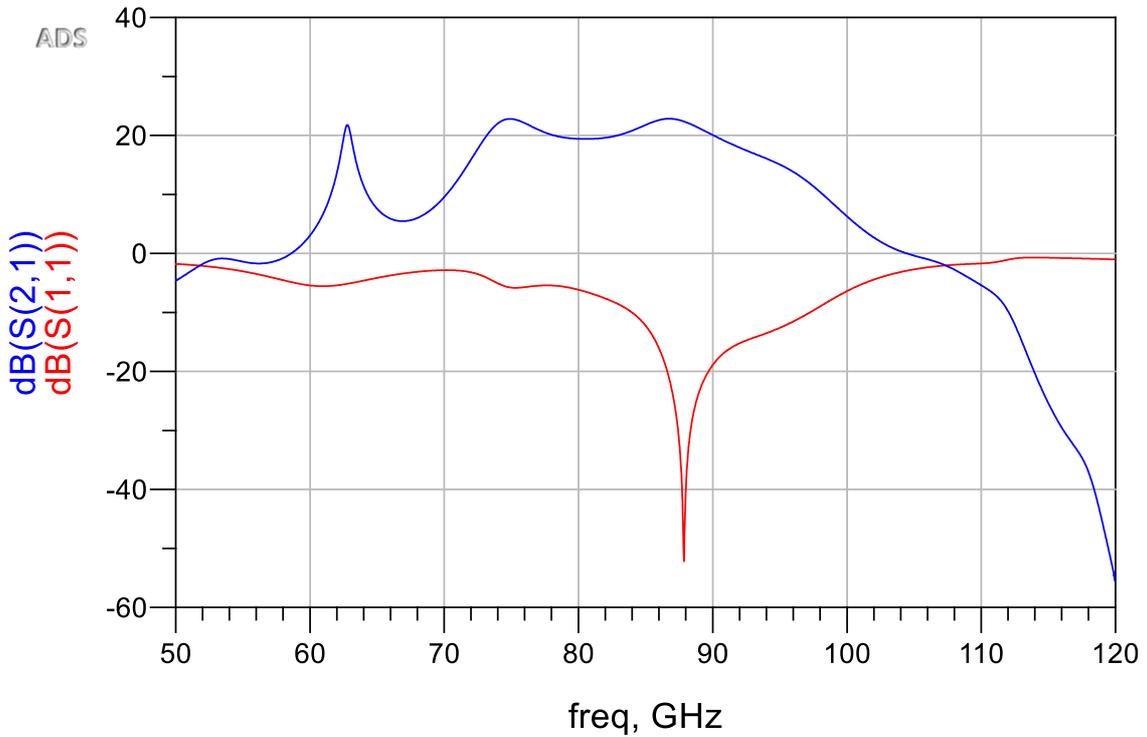

Figure 5. S-parameter analysis of the proposed PA.

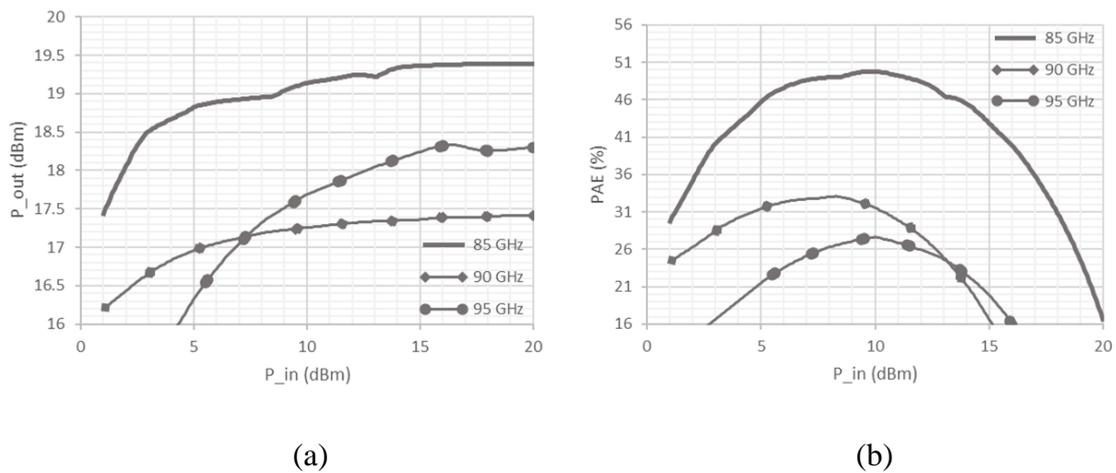

(a)                                              (b)

Figure 6. (a) PA linearity analysis and (b) PAE analysis at 85 GHz, 90 GHz, and 95 GHz.

The Monte Carlo analysis shows a close S-parameter result over 250 simulations, as depicted in Figure 7. The data analysis results are presented at Table 1 with 100 \% operation success in S11 (operate below - 10 dB) and S21 (operate above 15 dB); as well as 60.6 % (151 over 250 times) for S22 isolation (operate below - 10 dB).

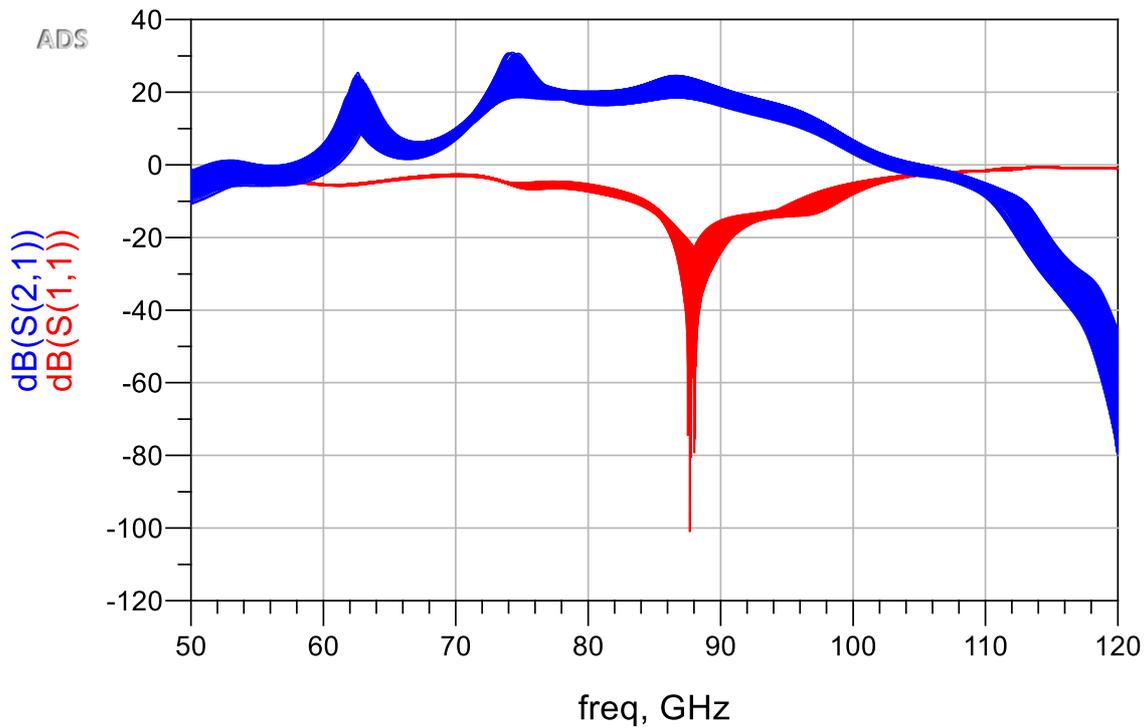

Figure 7. Monte Carlo analysis of the proposed PA.

Table 1. Frequency response over 250 times at 90 GHz.

| N=250 | S11 | S21 |
|---|---|---|
| **Min Value** | -24.29 dB | 15.77 dB |
| **Max Value** | -15.39 dB | 21.2 dB |
| **Standard Deviation, $\sigma$** | 2.297 | 1.358 |
| **Success Rate** | 100 % | 100% |

*3.1 Comparison with state-of-the-arts*

Table 2 summarizes the performance comparison of the designed PA with some of the state-of-the-art PAs covering sub-THz band and possessing potentials for future 6G transceiver applications. Overall, all the approaches use the common multi-staging and the power combining method for their PA designs. However, the performance of the PA varies due to transistor size, process, transistor $f_t/f_{max}$, stages, amplifier classes, source to drain voltages, biasing voltages, matching networks, layout designs, etc. The simulation results of this work show a very competitive performance, although using the largest process among all and is able to operate at W-band. The usage of high-power transistors from the foundry in the design contributed to achieve this performance trade-off.

Table 2. Performance comparison of the state-of-the-art PA designs.

| Ref. | Process | Topology | Power Combining | Freq. (GHz) | $BW_{3dB}$ (GHz) | Gain (dB) | $P_{sat}$ (dBm) | PAE (%) | $P_{DC}$ (mW) | Area ($mm^2$) |
|---|---|---|---|---|---|---|---|---|---|---|
| [28] | 28nm CMOS | 4 Diff. CS | 2-way | 99 | 10.3 | 21.8 | 15.1 | 18.6 | 32.4 | 0.054 |
| [29] | 65nm CMOS | 4 Diff. CS | 2-way | 109 | 17 | 20.3 | 15.2 | 10.3 | - | 0.34 |
| [30] | 40nm CMOS | 3 Diff. CS | 4-way | 120 | 38.5 | 16 | 14.6 | 9.4 | - | 0.33 |
| [26] | 100nm GaAs pHEMT | 3 Single-Ended CS | 2-way | 86 | 15 | 11.5 | 19.6 | 12.8 | 660 | 3 |
| [31] | 100nm GaAs pHEMT | 4 Single-Ended CS | 4-way | 88 | 19 | 15 | 22.3 | - | 1260 | 2 |
| [32] | 150nm GaAs pHEMT | 3 Single-Ended CS | 8-way | 33 | - | 21 | 33.6 | 25.6 | - | 18.3 |
| **This Work** | **150nm GaAs pHEMT** | **3 Single-Ended (2CS\*+1Cascode†)** | **2-way** | **90** | **21** | **20.1** | **17.2** | **33** | **72\*/ 186†** | **14.2** |

The gain of our design shows the highest value of 6.7 dB per stage. Although the design from [28] achieves the highest gain but it uses more stages to boost the gain values. The high gain value is achieved by using a cascode Class-A power stage having the ability to boost and deliver high power amplification [33]. The CS stage is able to provide an optimum low power performance; but this cannot bring any benefit at the power stage which is required to handle high power delivery from the previous stages, to boost the signal performance. However, the combination of CS and cascode stages can cause more design complexities in the layout, due to cascode connectivity with a larger size matching network as well as having a different biasing voltage than the CS driver stage.

Furthermore, the $P_{sat}$ of our design shows a competitive value with only 2-way power combining. The $P_{sat}$ performance is able to increase with a higher number of N-way power combiner. Although the design from [30] uses the most power combining of 4-way among all the other approaches, the $P_{sat}$ is the lowest. This can be explained with the wide bandwidth performance, and the small transistor size with higher parasitic that have degraded the $P_{sat}$ value. The usage of power divider is also an important factor that affect the $P_{sat}$ performance. The designs from [26, 31] used a Lange coupler for power combining which have a better $P_{sat}$ improvement than that of Wilkinson's power divider. A Lange coupler has a better phase balance, wide band and a high-power handing capability with a lower insertion loss [34]. Although Lange coupler can be more beneficial than Wilkinson's power divider, it has more design complexities especially for the smaller scale processes at higher frequencies.

Additionally, the PAE of this work shows the highest value among all the other reported designs in literature which is 33 % at 90 GHz. The most critical aspect that affects the PAE performance is the transistor size. When a smaller transistor is used in the design, at higher operating frequencies, the transistor can be seen to have a negligible impedance that easily increases the capacitive parasitic leakages. As our design uses the largest transistor size compared to other research, therefore, the highest PAE performance is achieved. On the other hand, the high $P_{sat}$ value achieved significantly improves the PAE as well [35]. However, transistors having large size can provide high PAE but the $f_t/f_{max}$ of the respective transistor will be lower, which makes it hard to operate at higher frequency. This implies that an optimized transistor size should be adopted to tune the operating frequency band and the PAE.

Lastly, the PA stages, in this work, are able to use a relatively smaller power of 72 mW and 186 mW for CS and cascode stages, respectively. This is done by carefully choosing the bias point during the DC-IV simulation. The PA design illustrated in [28] used the least power and can be easily explained as it used the smallest transistor size of 28 nm and can be easily biased with a smaller voltage. As this work uses the largest size of 150 nm (> 5 times larger than 28 nm designed compared to [28]), the overall PA layout design, including all the pads, is the largest too. Besides, the complex matching network required for the cascode stage contributed to the relatively larger layout. Furthermore, the authors of [32] used similar technology, however. our work shows a better overall performance tradeoff, particularly in terms of the operating frequency, PAE, and area compactness.

## 4. Conclusion

This research presented a multistage power amplifier design using a 150-nm GaAs technology for W-band application. The proposed PA utilize the benefit of CS driver stage and cascode power stage with capacitive biased operation, power combining and wide band matching network to mitigate the high parasitic at high frequencies. The designed PA has demonstrated a competitive performance especially for future 6G communication transceivers. The designed PA is able to operate at 90 GHz with a bandwidth of 20.7 GHz covering 72.3 to 93 GHz. The results shows that the proposed PA achieved the gain, $P_{sat}$, and PAE of 20.1 dB, 17.2 dBm, and 33 % respectively. It consumes a low power usage of 72 mW for CS stage and 186 mW for cascode stage for its operation. The layout design is also very compact which is $5.66 \times 2.51$ mm$^2$ including all the pads. Our PA circuit will significantly improve the integrated circuits based designs at sub-THz which in turn would lead to widespread adoption of the emerging 6G communication technology.

**Nomenclature**

| | |
|---|---|
| $BW_{3dB}$ | 3-dB bandwidth |
| CE | Common emitter |
| CG | Common gate |
| CS | Common source |
| CMOS | Complementary metal-oxide-semiconductor |
| DC-IV | Direct-Current Current-Voltage |
| $f_t/f_{max}$ | Transit frequency/maximum oscillation frequency |
| GaAs | Gallium arsenide |
| ML | Microstrip lines |
| mmW | Millimeter-wave |
| PA | Power amplifier |
| pHEMT | Pseudomorphic high electron mobility transistor |
| $P_{sat}$ | Saturated output power |
| PAE | Power added efficiency |
| Q | Matching locci |
| THz | Terahertz |
| TL | Transmission lines |
| 5G | Fifth-generation |
| 6G | Sixth-generation |

**Acknowledgement**

The authors would like to thanks United Monolithic Semiconductors (UMS) for providing the PDK file to support the research. This research is financially supported by Xiamen University Malaysia (Project code: XMUMRF/2021-C8/IECE/0021). The authors would like to thank the reviewers' constructive suggestions and comments.

**Conflicts of interests/Competing interests**

The authors declare no conflicts of interests/competing interests.